\documentstyle[12pt]{article}
  \evensidemargin=\oddsidemargin
\def\'#1{\ifx#1i{\accent"13\i}\else{\accent"13#1}\fi}
\def\ref{\par\noindent\hangindent=2pc \hangafter=1 }
\def\B{$B$}
\def\K{$K$}
\begin{document}
\pagestyle{plain}
\title{\LARGE \bf Number and Luminosity Evolution of Interacting
Galaxies as a Natural Explanation for the Galaxy Counts}
\author{Pedro Col\'in}
\date{}
\maketitle

\centerline{Instituto de Astronom\'ia, Universidad
Nacional Autonoma de M\'exico}
\centerline{CP 04510 M\'exico, D.F. M\'exico}
\centerline{e-mail:colin@astroscu.unam.mx}
\centerline{{\it accepted for publication in} RevMexAA}

\begin{abstract}
A newly developed isochrone
synthesis algorithm for the photometric
evolution of galaxies is described. Two initial mass functions, IMFs,
in particular, the recent IMF determined
by Kroupa, Tout, and Gilmore, three photometric transformations,
and a 1-Gyr-burst star formation rate, SFR, are used to compute the $B-V$ and
$V-K$ color index evolution.  Non-negligible differences are observed
among model results.

In the framework of the galaxy count model by
Col\'in, Schramm, and Peimbert a simple merging scenario is considered to
account for the excess of galaxies observed in the blue band counts. The
excess is explained by the number and luminosity evolution of a group of
galaxies called interacting, I. It is
assumed that the number of I galaxies increases
as $(1+z)^{\eta}$ due to mergers.
Moreover, it is proposed that their
characteristic luminosity increases
as $(1+z)^3$ due to
starbursts driven by galaxy-galaxy collision and decreases as $(1+z)^{-\eta}$
due to the change in the size of the galaxies.
Not much number evolution is needed
to account for the excess; for example, a
model with $\eta = 4.0$ predicts that about 17 \%
of the galaxies at $z = 0.4$ are interacting.
Number evolution models with a rather high value of $\eta$ fit better
the data; in particular, the model with $\eta = 4.0$ predicts that about
13 \% of the galaxies have $z > 0.7$
in the $21.0 < m_{b_J} < 22.5$
interval, this contrasts with the upper bound of 5 \% obtained with
the sample of 78
galaxies by Colless et al. The excess of high redshift galaxies can
not be simply explained by changing reasonably the parameters of the
luminosity function
of I galaxies. This result could indicate that mergers are not the
whole story.
Our best-fit model produces the following values for
the parameters of the local luminosity function of galaxies:
$\alpha = -1.20$, $M^* = -20.7$, and $ \phi^* = 1.66 \times 10^{-3} Mpc^{-3}$
($h =0.5$).
\end{abstract}

\section{INTRODUCTION}
The excess number density of galaxies observed in the blue
band (e.g. Tyson 1988; Metcalfe et al. 1991;
Lilly, Cowie, \& Gardner 1991) over the expected
from a non-evolving model has generated
great interest. Several possible explanations
have been suggested to account for it: from those
that invoke a nonzero cosmological constant that provides a greater
volume to accommodate the excess number of galaxies
(Yoshii \& Takahara 1988; Fukugita et al. 1990; Yoshii 1993), to those
that assume evolution of the luminosity function
of galaxies (LFG) via an increase in the number density of
galaxies due to mergers (Guiderdoni \& Rocca-Volmerange 1987;
Broadhurst, Ellis \& Glazebrook 1992; Col\'in, Schramm, \& Peimbert 1994,
hereafter CSP)
or via an increase in the characteristic luminosity of the LFG due to
starbursts driven possibly by galaxy-galaxy
collisions (Carlberg \& Charlot 1992). Moreover,
faint redshift surveys by Broadhurst et al. (1988) and
Colless et al. (1990, 1993) place galaxies quite near,
the median redshift is just $\sim 0.2$ and $\sim 0.3$, respectively.
In addition,
no galaxies were found with $z > 0.7$ brighter than $ b_J = 22.5$
in the sample by Colless et al. This result
imposes a strong constraint on evolving models. In
these models a non-negligible
fraction of galaxies with $z > 0.7$ is expected, indeed larger than
the one predicted by a non-evolving model.

  With the advent of infrared arrays and infrared CCD
detectors it is also possible to count galaxies deeply in the
\K-band (Cowie et al. 1993; Soifer et al. 1994). Surprisingly,
a non-evolving model fits relatively well the \K-counts. This
almost rules out models with nonstandard cosmologies,
as the ones with a non-zero cosmological constant, since
in this case they overestimate the number of galaxies.
Yet, the observational restrictions
admit one more explanation, namely a new population of blue
galaxies at $z > 0.2$ that faded or were disrupted so as to
become invisible at present (Cowie, Songaila, \& Hu 1991; Babul \& Rees 1992).

Local low surface brightness galaxies have similar physical
properties to those observed in the
faint blue excess population (McGaugh 1994). McGaugh suggests
that the excess is made up of low surface brightness galaxies
and his idea rests on a bivariate LFG; that is, a LFG that, in
addition to type, depends on surface brightness.

The result that faint blue galaxies (FBGs) are weakly clustered
(Efstathiou et al. 1991)
has been sometimes taken as an evidence against the merger-driven
explanation for the observed excess of FBGs (see, for
instance, Babul \& Rees 1992;
McGaugh 1994);
the low value for the angular correlation function at
$30''$ is explained by Efstathiou et al. assuming that most
FBGs belong to a population that is weakly clustered and intrinsically
faint at the present epoch. This result is challenged by a
recent work by Cole et al. (1994) where no evidence is found
for the evolution of the comoving correlation length
with redshift to $B= 22$, however a more recent paper by Infante \& Pritchet
(1994) supports the earlier results by Efstathiou et al.

The hypothesis that the observed excess of FBGs
is due to an increase in the number density of galaxies that are
strongly interacting, the merger-driven excess, is
retaken here.
A phenomenological model
was developed in CSP to account for three different observational
restrictions, namely: (1) the \B-counts, (2) the \K-counts,
and (3) a \B-redshift distribution. It was due to the
increase in the number density of a type of galaxies, called interacting (I),
that the observations were fitted.

This kind of model, like others of similar nature,
is supported by a recent work on high-resolution
imaging of faint blue galaxies (Colless et al. 1994)
and by
HST studies of morphology of high redshift galaxies (Griffiths et al.
1994a, b).
Colless et al. find, from a sample of 26 galaxies which
belong to the redshift survey of Colless et al. (1993), 17 galaxies
that have an enhanced star formation rate indicated by
[OII] equivalent widths greater than 20\AA. About 30\% of these
galaxies have companions at projected distances closer
than 10 h$^{-1}$ kpc.

This model differs from the one by CSP in the following: (1) a
different algorithm for the computation of the `passive' luminosity evolution
correction is applied, and (2) although the total luminosity of the I galaxies,
$L_I^T$, here is also not
conserved, it is not due to that we do not take into account the change in
size of the I galaxies, as it occurs in CSP, but
due to the starsbursts driven by galaxy-galaxy collisions. This is
what we call `active' luminosity evolution. The model in this paper
also differs from the one by CSP in the assumed values for the
parameters of the luminosity functions of the different types of galaxies; in
particular, those values assumed for the characteristic luminosities of Sa--c
and E/SO galaxies. Because they are fainter in CSP, they produce a better fit
for the redshift distribution than the one we show in this paper.

An outline of the paper follows: in \S2 we briefly discuss
our revised color evolution code with an isochrone synthesis
algorithm, in \S3 our merger-driven model
is described, and in \S4 we present our conclusions.

\section{COLOR EVOLUTION OF GALAXIES}

\subsection{Model}

Our photometric evolution of galaxies code is fully discussed
in CSP therefore here just a brief review
will be given. Unlike CSP where a standard procedure
was used (e.g. Bruzual 1983) here an isochrone synthesis
algorithm is implemented. The advantages of this algorithm have
been discussed elsewhere (e.g. Charlot \& Bruzual 1991), we just
would like to mention the usefulness of the algorithm in computing
color evolution properties for short ($\sim 10^8$ yr) time-scales
of the star formation rate.

\subsubsection{Ingredients}

The ingredients necessary in any spectrophotometric
evolution of galaxies code are four: (1) a library of stellar evolutionary
tracks, (2) a library of stellar spectra, (3) an initial mass
function (IMF), and (4) a star formation rate (SFR). Our compiled
tracks are from Schaller et al. (1992) for solar metallicity. These
tracks are incomplete and therefore we had
to complete them. Tracks from other authors were used and
various useful simplifications were made; in particular,
the lifetime of the asymptotic giant branch was left as a free
parameter. This will be remedied as soon as the evolutionary models
for the horizontal and asymptotic giant branch from the group
of Maeder and collaborators are incorporated.
As long as we are only
interested in galaxy color evolution there is no need for
having a library of stellar spectra, we just need a photometric
transformation from the HR theoretical diagram to the observational
one. The compilation work by Schmidt-Kaler (1982) has been used now
to pass from one diagram to the other. We still continue to use the
work by Johnson (1966) to get the rest of the color indexes.

The IMF is normalized so that,
\begin{equation}
\int_{m_{low}}^{m_{up}} m \phi(m)dm = 1.
\end{equation}
Although a value of 0.1 $M_\odot$ is often used
for $m_{low}$, as we do, we should keep in mind
that the contribution by substellar objects to
the integrated infrared light of a galaxy
might not be negligible. Of course, this
depends on the luminosity function of these
objects of which we know nothing. From chemical
evolution considerations Peimbert et al. (1994)
limit the amount of substellar objects in the
solar neighborhood to $<$ 0.02 $M_\odot pc^{-3}$.
The recent IMF determination by Kroupa, Tout, and Gilmore (1993), KTG IMF,
for the solar vicinity is used, as a good guess,
for the IMF for a variety of types of galaxies. Nevertheless,
the ``standard'' Salpeter IMF, S IMF, with the exponent $x=1.35$
is used for comparison purposes. A value of 60.0 $M_\odot$
is taken for $m_{up}$.

In view of the great uncertainties about
stellar formation histories of galaxies,
an exponential law for the SFR is usually
taken to model
the color evolution of the different types of galaxies.
The maximum virtue of this scheme is that almost all present galaxy
colors can be reproduced by varying the time scale, $\tau$,
of the SFR: from an elliptical galaxy, $\tau = 0.5$, to an
irregular one, $\tau = \infty$ (e.g. Charlot \& Bruzual 1991; CSP).
These galaxy colors are, of course, metallicity-dependent.
This effect is now being incorporated in the latest
models for spectral evolution of galaxies; in particular, for
elliptical galaxies see Worthey (1994 and references
therein). The evolutionary correction, e-correction, to the counts
should not worry us too much because as
we know from CSP there are only three types of galaxies that
play an important role on the counts.
According to our count-scheme, the luminosity
of a galaxy belonging to the interacting group of galaxies does
not evolve; i.e.,
it has no e-correction. We
can be confident that the other two types, E/SO and Sa-c,
are being modeled relatively well; the first,
with a burst of star formation that lasts 1 Gyr, the second,
with a constant SFR.

\subsubsection{Isochrone synthesis algorithm}

The luminosity of a single stellar population (SSP) is given by
\begin{equation}
l_{\Delta \lambda} (t)= \int_{m_{min}}^{m_{max}}
10^{-0.4(M_{\Delta \lambda}-M_{\Delta \lambda, \odot})} \phi (m) dm,
\end{equation}
where $m_{min}$ is the lower limit of the IMF and $m_{max}$
is the maximum mass in the isochrone with age $t$.  The integrated luminosity
for a composite stellar population, with a history of star formation given
by $\psi (t)$, can be obtained with the convolution integral
\begin{equation}
L_{\Delta \lambda} (t)= \int_0^t \psi (t-t') l_{\Delta \lambda} (t') dt'.
\end{equation}
The isochrone at time $t$ from a complete set of m
evolutionary tracks is built
as follows: let us assume that the tracks are divided into n stages, from zero
age main sequence (ZAMS) to post-asymptotic
giant branch (P-AGB). The indexes $i$
and $j$ number the stages and the tracks, respectively, with
$i= 1,...,{\rm n}$ and
$j= 1,...,{\rm m}$. The mass of the star at the stage $i$ is given by
\begin{equation}
\log m_i (t)= A_{i,j} \log (M_{j+1}) + (1-A_{i,j}) \log (M_j),
\end{equation}
where
\begin{equation}
A_{i,j}= {\log t_{j,i} - \log t \over \log t_{j,i} - \log t_{j+1,i}}.
\end{equation}
In equation (5) $t_{j,i}$ denotes the age of the star of ZAMS-mass $M_j$
at the $i^{th}$ evolutionary stage, where
\[t_{j+1,i} \le t < t_{j,i},\]
and
\[M_j \le m_i (t) < M_{j+1}.\]
$M_1$ and $M_{\rm m}$ denote the  lower and upper limit of the IMF,
respectively. This procedure is
performed for each stage (value of $i$).  The HR diagram physical parameters,
$\log L$ and $\log T_{eff}$, associated to the star of mass $m_i$
are obtained by interpolating the tracks between the stars of masses
$M_j$ and $M_{j+1}$.

To obtain the integrated properties of this SSP we proceed as
Bruzual (1992): the number of stars of
mass $m_i$, $n(m_i)$, is found by integrating the IMF
from $m^-$ to $m^+$, where $m^-=
\sqrt {m_{i-1} m_i}$ and $m^+ = \sqrt {m_i m_{i+1}}$. The
luminosity, $l_{\Delta \lambda, i}$, is then computed by
assigning $N_{i,j}= A_{i,j}n(m_i)$ stars to the HR parameters of
the star of mass
$M_{j+1}$ at the $i^{th}$ stage, and $N_{i,j+1}= (1-A_{i,j})n(m_i)$
to the star of mass $M_j$ at the same $i^{th}$ stage. We finally
obtain $l_{\Delta \lambda} (t)$ by summing over all $i's$.

\begin{figure}
        \vspace{12cm}
        \includegraphics{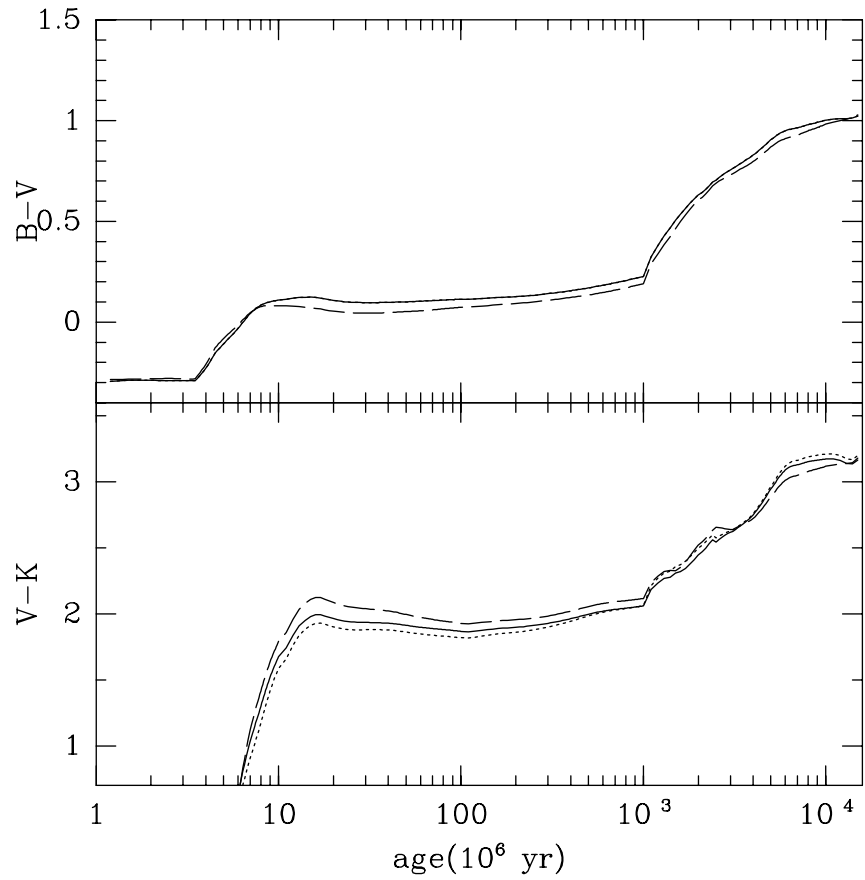}
        \caption{Evolution of the color indices B-V and V-K for
a 1Gyr-burst SFR with three different photometric calibrations:
Bruzal \& Charlot (dashed line), Schmidth-Kaler and Bessel \& Brett
(solid line), and Schmidth-Kaler and Johnson (dotted line).}
\end{figure}

\subsection{Results}

In doing photometric evolution of galaxies special
care should be put on the photometric transformation. This
is stressed in Figure 1 where the B-V and
V-K evolution is plotted for a 1-Gyr-burst SFR,
for three different calibrations.
The dotted
line uses a compilation work by Schmidt-Kaler (1982) and Johnson (1966),
the solid line utilizes one from Schmidt-Kaler (1982) and Bessel
and Brett (1988), and the dashed line comes from a
compiled calibration work by Bruzual and Charlot (1993). The
greatest difference among them
amounts to $\sim 0.25$ mag in V-K at around $1.7 \times 10^7 yr$;
this difference is due to the V-K values adopted for the giant
luminosity class
at the low-temperature regime. The peak at $1.7 \times 10^7 yr$
is produced by the supergiant red branch contribution. Hereafter
the calibration by Schmidt-Kaler (1982) and Bessel and Brett (1988)
will be used. In Figure 2 we have plotted the B-V and V-K color
evolution for a 1-Gyr-burst SFR, in this case for two different IMFs:
the KTG IMF (solid line) and the Salpeter one (dotted line).
The V-K color is slightly bluer when
computed with the KTG IMF, the greatest difference at $\sim 10^8yr$
amounts to 0.1 mag. A much smaller difference is observed for the
case of B-V.

\begin{figure}
        \vspace{13cm}
        \includegraphics{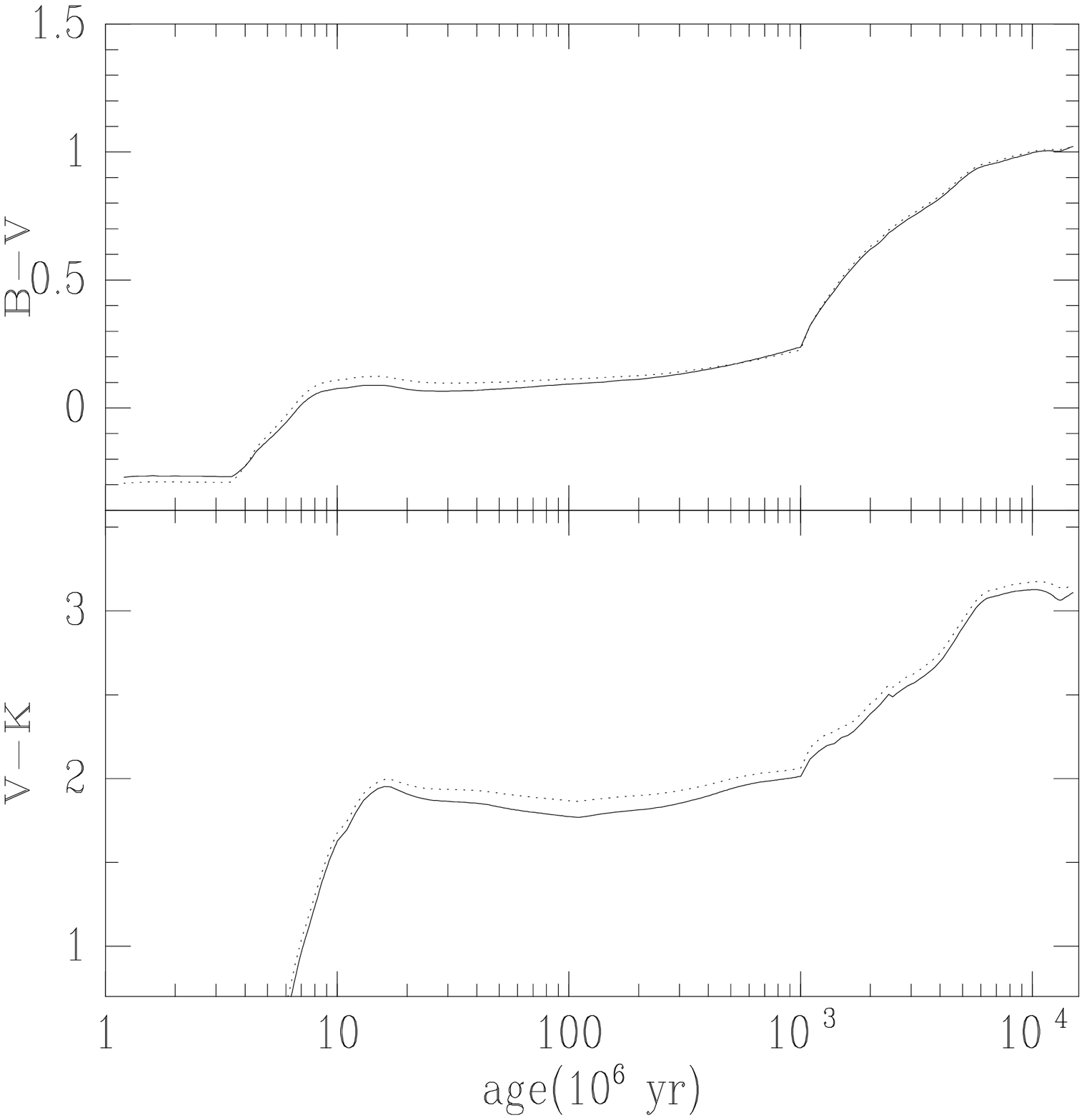}
        \caption{As in Fig. 1 we have plotted
the evolution of the color indices B-V and V-K for
a 1-Gyr-burst SFR. The two curves come from two different IMFs: KTG IMF
(solid line) and Salpeter IMF (dotted line). The calibration from
Smichdt-Kaler and Bessel \& Brett has been used.}
\end{figure}

\section{A Merger Model for Galaxy Counts}

This paragraph describes briefly the $\Omega_0 = 1$ merger model for galaxy
counts of CSP. Galaxies were divided into five different types:
E/SO, dE/dSO (spheroidal dwarfs), Sa-c (spirals), dI (dwarf irregulars),
and I (interacting galaxies). A Schechter analytical representation for
the galaxy luminosity function was used for all types.
It was shown that a gaussian representation for the
luminosity function (LF) of the brighter galaxies provided a poor fit
to the redshift distribution. CSP models were based on
a scenario where the amplitude of the LF of I galaxies,
$\phi^*_I$, increased with redshift as a power law; i.e.,
$\phi^*_I \propto (1+z)^\eta$. The k-corrections
for the different classes of galaxies were obtained from the
spectral energy distributions (SEDs) by Coleman, Wu, \& Weedman (1980)
and Pence (1976); in particular, the SED for an average
elliptical galaxy was taken to be the SED of the bulge of M31. The
e-corrections were evaluated from our standard approach to the
photometric evolution of galaxies.

\subsection{The Interacting Galaxies}

The I group does not form properly a morphological group of galaxies,
it is rather composed of pairs of strongly interacting galaxies of
different nature. Moreover, the
observed physical characteristics of the faint galaxies such as: their
blue colors, enhanced star formation, indicated by [OII] equivalent
widths greater than 20 \AA\ (Colless et al. 1994), etc. give us
information about the nature of the
galaxies that intervene in the interactions.
These
characteristics, which are often associated to
a system of interacting galaxies, seem to indicate that the type of galaxies
that are participating in the mergers are gas-rich galaxies
(e.g. Carlberg \& Charlot 1992). This is a point worth noting
since we will see below that
the increase of the characteristic luminosity of the I galaxies, $L^*_I$,
to a first approximation, is due to the starbursts driven by
cloud-cloud collision. It will be assumed
that the I-galaxy LF (LFI) follows a Schechter form.

The evolution of LFI is due to three factors:
(1) a greater number density,
(2) a greater gas content that results in a
stronger starburst, and (3) a smaller average size.
The points (2) and (3) go in opposite direction, whereas the point (2)
increases $L^*_I$ the point (3) decreases it. It is interesting to note
that in the absence of starbursts
the effects on \B-counts by points (1) and (3) cancels out. This might
be the case for the \K-counts where most of the light
comes from a population of old stars.

Our merger model scenario for the galaxy counts is then the
following: suppose that at each time a photograph of the Universe is taken
and that galaxies are divided into various types. Those that show signs
of strong interaction are brought to the I group of galaxies.
As we look back in time we see that the number density (per comoving
volume) of I galaxies
increases, while, by hypothesis, the number density of the other types of
galaxies does not change. The increase in the total luminosity of I galaxies
($\propto L^*_I \phi^*_I$) is due to the
their greater amount of gas as well as their increasing number density.

In this paper the above scenario for the modeling
of the \B-, \K-counts, and the redshift distribution
is used and compared with observations.
We propose that the $L^*_I$ of both the \B-band and
the \K-band evolves due to galaxy-galaxy collision and due to the change
in size, yet a model for the \K-counts where the contribution
from starbursts is neglected is also computed.
The galaxy-galaxy collision term,
the one which produces an `active'
luminosity evolution, contributes
with $(1+z)^3$ to $L^*_I$ and it comes from
assuming that: (a) $L^*_I$ is proportional to the stellar formation, $\Psi$,
driven by the mergers, (b) in a cloud-cloud
collision $\Psi \propto \sigma^2$, where $\sigma$ is the gas density,
and (c) $\Omega_0 = 1$.
On the other hand, $L^*_I$ is reduced by $(1+z)^{-\eta}$ due to
the change in size.
Models with different values of $\eta$ are calculated.

A Schechter analytical representation is used for
the LFI,
with relative freedom in choosing its $\alpha$ and $M^*$ parameters
due to the great uncertainty about its shape.
Its characteristic luminosity, the luminosity at the ``knee''
of the LF, is assumed
to be greater than the corresponding value for elliptical and
spiral galaxies. The fraction of I galaxies that contributes
to the LFG is assumed greater than the
amount of mergers observed at the present epoch (1--2 \%) in
order to improve our fit to the $B$-counts.

\subsection{Ingredients}

In Table 1 the values of the parameters of the luminosity functions
of the different galaxy classes are shown:
second, third, and fourth columns. In the
fifth, sixth, an seventh columns the timescale of the SFR, the $B-V$
and $B-K$ color indices are shown, respectively. These later are
computed using the
photometric evolution code described in \S2. The last column indicates the
kind of SED used for each type of galaxy; in particular, the SED of the
bulge of M31 is used for E/SO galaxies.

The assumed values for $\alpha$, $M^*$, and the mix reflect several facts.
First,
the values for the parameters $\alpha$ and $M^*$ of the LF of spiral galaxies
are similar to those derived by Loveday et al. (1992) for the field LFG.
Second,
there is still much uncertainty about the value of $\alpha$ for
early-type galaxies,
the rather low value assumed in this paper, as compared with those
obtained by several authors
(Efstathiou et al. 1988; Loveday 1992; Zucca, Pozzetti, \& Zamorani 1994),
fits better the \K-counts. Third, there is not yet a field LF for early-type
dwarf
galaxies nor for late-type dwarfs, though we use the $\alpha$ and $M^*$ values
of the LF of dE/dSO galaxies and the $M^*$ value of the LF of
dI galaxies obtained by Sandage, Binggeli, \& Tammann (1985)
for the Virgo cluster; needless to say that there is no reason
to expect that these
assumptions are correct.  Fourth, the parameters for the LFI are chosen
in such a way that: (1) the characteristic luminosity
is greater than the one for
E/SO galaxies, (2) the slope at the faint end equals that of the total
luminosity function, and (3) the fraction of I galaxies should be small.
At the end, our mix takes into account the
following facts: (1) the uncertainty in the fraction of strongly
interacting systems,
(2) there are more spiral galaxies than
elliptical galaxies in the
surveys of Efstathiou et al. and Loveday et al., (3)
there is a significant percentage
of dE/dSO galaxies in clusters (Ferguson \& Sandage 1991), in the
poorest cluster,
the Leo group, about 46 \% are dE/dSO galaxies, and (4)
recent determinations of the LFG (Eales 1993; Lonsdale \& Chokshi 1993)
are producing a steeper slope
at the faint end ($\sim -1.3$).

\begin{center}
\centerline {TABLE 1}
\vspace{0.3cm}
\centerline{INPUT PARAMATERS OF THE MODELS}
\vspace{0.2cm}
\begin{tabular}{ccrlcccc}\hline\hline
 & & & & $\tau^a$ & & & \\
Type & $\alpha$ & $M^*$ & Mix &($Gyr$) & $B-V^b$ & $B-K^b$ & SED$^c$ \\
\hline
I & $-1.20$ & $-21.2$ & 0.05 & $\infty$ & 0.49 & 3.04 & Sdm \\
dI & $-1.40$ & $-16.2$ & 0.16 & $\infty$ & 0.49 & 3.04 & Sdm \\
Sa-c & $-0.80$ & $-20.0$ & 0.30 & $5$ & 0.70 & 3.54 & Scd \\
dE/dSO & $-1.35 $ & $-18.0$ & 0.30 & $5.0$ & 0.70 & 3.54 & Sab \\
E/SO & $-1.00 $ & $-20.7$ & 0.19 & $0.5$ & 1.01 & 4.17 & M31 \\ \hline
\end{tabular}
\end{center}
{\footnotesize
\noindent $^a$ The star formation rate used to get the color of the galaxies at
the present epoch is: $\psi \propto e^{-t/\tau}$.

\noindent $^b$ Two local color indices representatives of the
different
types of galaxies,
computed by our photometric evolution of galaxies code.

\noindent $^c$ The spectral energy distributions, SEDs, are from CWW
and Pence 1976; in
particular, for E/SO galaxies we have taken the SED of the bulge of M31.}
\vspace{0.7cm}

The e-corrections are evaluated assuming
an exponential decreasing SFR with a varying timescale.
For dwarf galaxies, spheroidal and irregulars,
it appears inappropriate to model their light
with a solar metallicity population. Furthermore,
a choice for the SFR becomes difficult; dwarf
irregular galaxies are better modeled by a series
of star formation bursts followed
by periods of zero star formation (e.g. Pilyugin 1993).
The number and the strength
of the bursts, and the duration of the quiescent periods are
parameters that seem to depend on the galaxy. Fortunately,
these galaxies make a negligible contribution to
the \B-, and \K-counts except for the fainter end (e.g. CSP).
The I galaxies have no e-correction by hypothesis.

\subsection{Results}

The detection rate
and selection effects, according to Yoshii (1993), were evaluated
but they were
not incorporated to our models. The reason is the
small effect they have on the predicted \B-counts. In
an isophotal magnitude scheme, $S_L= 29$ \B\ mag arcsec$^{-2}$, with an
imposed minimum diameter of $D_{min}= 2.0''$, a detection
rate of about 75\% at $B = 26$ is expected in a
$(\Omega_0,\Lambda_0)=
(1,0)$ scenario. This amounts to
a difference of just 0.10 dex at $B = 26$ (see also Yoshii).

In Figure 3 the number of galaxies versus blue magnitude is plotted
for four models. The data are from Maddox et al. (1990), Lilly et al. (1991)
and Metcalfe et al. (1991).
There are two non-evolving, NE, models: one with
the parameter values from Table 1 (NE) and the other one
(NE KGB) with  the parameter values from
Koo,  Gronwall, \& Bruzual (1993, hereafter KGB). The Table 2
from KGB was utilized to derive the Schechter parameters for
the I ($B-V < 0.6$), Sa-c ($0.6 \le B-V < 0.85$), and E/SO ($B-V \ge 0.85$)
classes of galaxies, these are:
\[ (\alpha,M^*,\phi^*)= \left\{ \begin{array}{ll}
    (-1.86,-21.6,9.93 \times 10^{-5}) & \quad \mbox{for I} \\
    (-1.13,-21.8,3.29 \times 10^{-4}) & \quad \mbox{for Sa-c} \\
    (-1.38,-20.1,9.90 \times 10^{-4}) & \quad \mbox{for E/SO}
\end{array} \right. \]
The inclusion of the NE KGB model in Figure 3
is with the purpose of showing that a much smaller difference is obtained
between a NE model and the observations if the parameters of the LFs
are moved in the `correct' direction.
The other two models plotted in this Figure are merger-driven
models for two values of $\eta$. A significant
difference is starting to
observe between each
other at $b_j > 25$. Moreover, the maximum difference between
the data and our $\eta = 4.0$ model amounts just 0.2 dex at $b_J = 25.0$.

\begin{figure}
        \vspace{12cm}
        \includegraphics{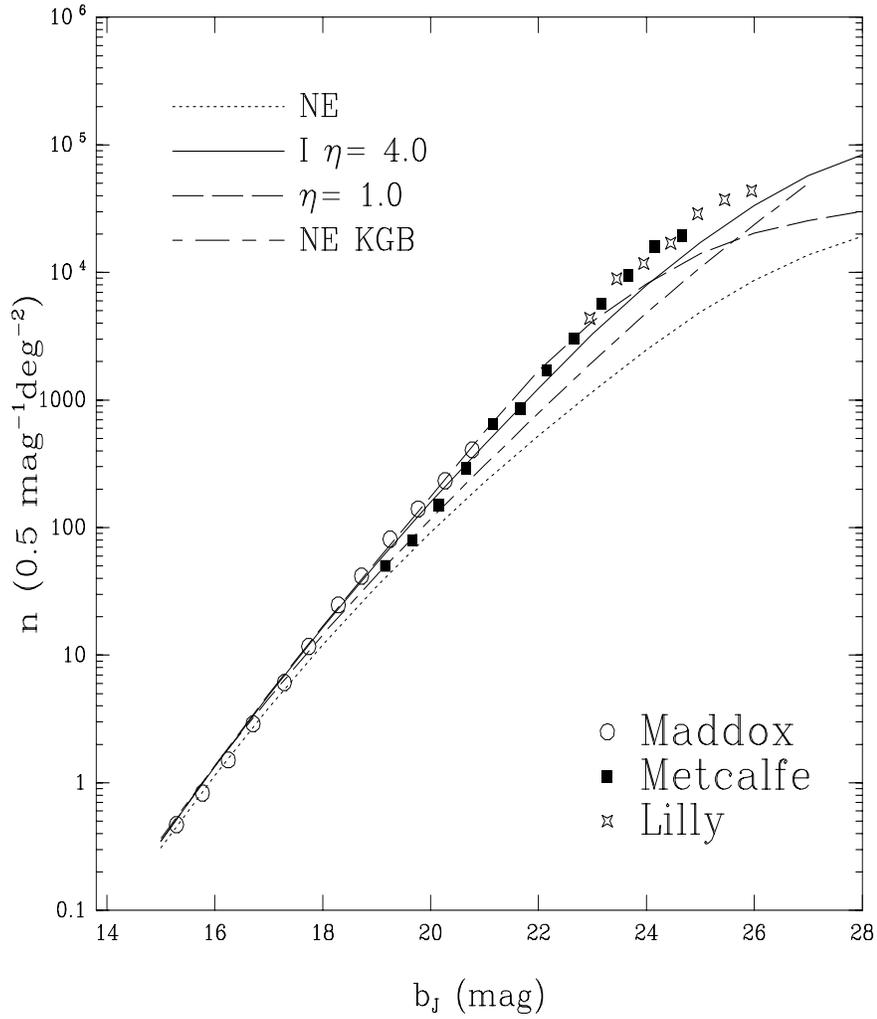}
        \caption{The number of galaxies versus blue magnitude is
plotted for four models: two are non-evolving, NE, and
the other two are merger-driven, for two values of $\eta$.}
\end{figure}

When the LFG is parameterized by a Schechter function, the number
density, $n_0$, is related
to the amplitude of the LFG, $\phi^*$, by
\[
n_0= \phi_* \Gamma (1+\alpha,\beta),
\]
where $\Gamma$ is the incomplete
gamma function and $\beta$ is the faint luminosity limit in units of $L^*$.
Our LFG is a composite LF, the sum of five different Schechter LFs,
therefore the parameters $\alpha$, $\phi^*$, and $M^*$
are not well defined. Yet, approximate values
for $\alpha$, $\phi^*$, and $M^*$ can be obtained by attempting to fit
our LFG to a Schechter function. When this is done, (see Figure
4) we get the following: $\alpha= -1.20$, $M^* = -20.7$, and $\phi^*=
1.66 \times 10^{-3} Mpc^{-3}$. It is encouraging to note
that the value we get for $\phi^*$, necessary to fit the brighter part
of the \B-counts, is similar to those derived by several authors (Efstathiou
1988;
de Lapparent, Geller, \& Huchra 1989; Loveday et al. 1992).

\begin{figure}
        \vspace{9.5cm}
        \includegraphics{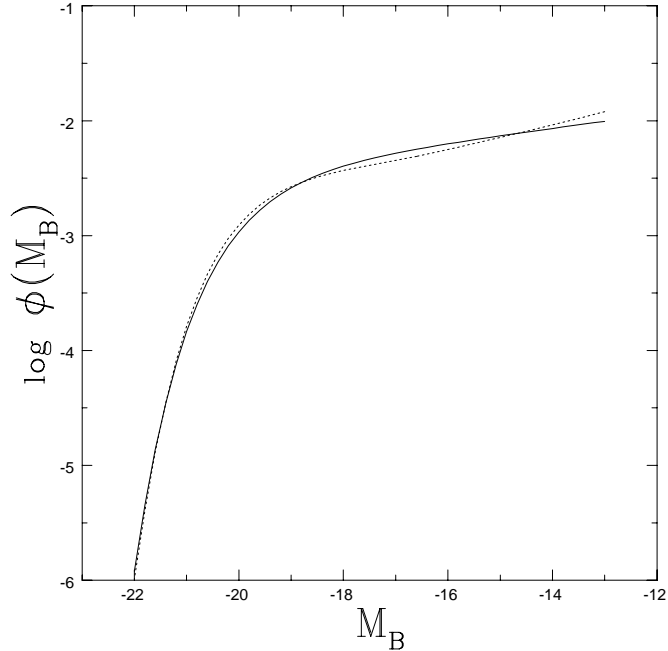}
        \caption{The logarithm of the LFG is plotted here (dotted line) along
with its Schechter fit (solid line). The values for the
parameters $\alpha$, $\phi^*$, and $M^*$
are given in the text.}
\end{figure}

In Figure 5, the number distribution in the \K-band is plotted for the
same models as for the \B-band except that here, the NE KGB model
is replaced by a merger-driven model in which the $L^*_I$ in the \K-band
does not increase due to the startbursts (model denoted by II $\eta = 4.0$).
The data are from a compilation by Gardner,
Cowie, \& Wainscoat (1993). It is important to point out here the better
agreement obtained by those models with a rather high value of
$\eta$. A redshift distribution is plotted in Figure
6 for two intervals of \B\ magnitude. The histograms are redshifts
for 125 galaxies taken from Broadhurst et al. (1988) in the $20.0 \le
b_J \le 21.0$
range and for 78 galaxies taken from Colless et al. (1990, 1993)
in the $21.0 \le b_J \le 22.5$ interval, this later survey is complete
to 4.5 \%. Our merger-driven models predict a excess of high redshift
galaxies, being it greater for models with a lower value
of $\eta$: the model with $\eta= 1.0$ predicts that a 27 \% of the galaxies
have
$z > 0.7$ in the  $21.0 \le b_J \le 22.5$ range,
clearly inconsistent with the 5 \% derived by Colless et al.
This is interesting since it could be a manner to discriminate
between models with number luminosity from those with luminosity
evolution. The model with $\eta = 4.0$
predicts that about 13 \% of galaxies have $z > 0.7$, not too
far away from the observed 5 \%.

It appears difficult to reduce the number of high redshift
galaxies by
changing the parameters of the LF of interacting galaxies, subject to
the constraints that: (1) a small percentage ($< 5$\%) of galaxies are
strongly interacting at the present time, (2) the slope at the faint end
is that of the total luminosity function, and (3) the characteristic
luminosity of this population, $L_I^*$,
is greater than the one of E/SO
galaxies.  For the sake of the argument, let us assume we increase
further the value of the parameter $\eta$. One
might think this would give us a better fit, but the point is that
the percentage of galaxies greater than 0.7
does not decrease much by increasing the value of $\eta$;
for example, for $\eta = 10$ it
is just 12\% compared to the 13\% for $\eta = 4$!. This point
can be explained as follows: as the $\eta$ value goes up the
$L_I^*$ decreases as
$$
M_I^* = M_I^*(0)-2.5(3-\eta)log(1+z),
$$
so the number of high-z galaxies decreases (improving the fit) but
at the same time the number of moderate z galaxies increases due to
the $(1+z)^\eta$ law (worsening the fit). At the end, the decreasing of
the percentage of galaxies with $z > 0.7$ is very small.
We took the value of $\eta = 4$ as a resonable estimation of the amount of
mergers at $z=0.4$ and as a not too bad fit to the
z-distribution.

\begin{figure}
        \vspace{12cm}
        \includegraphics{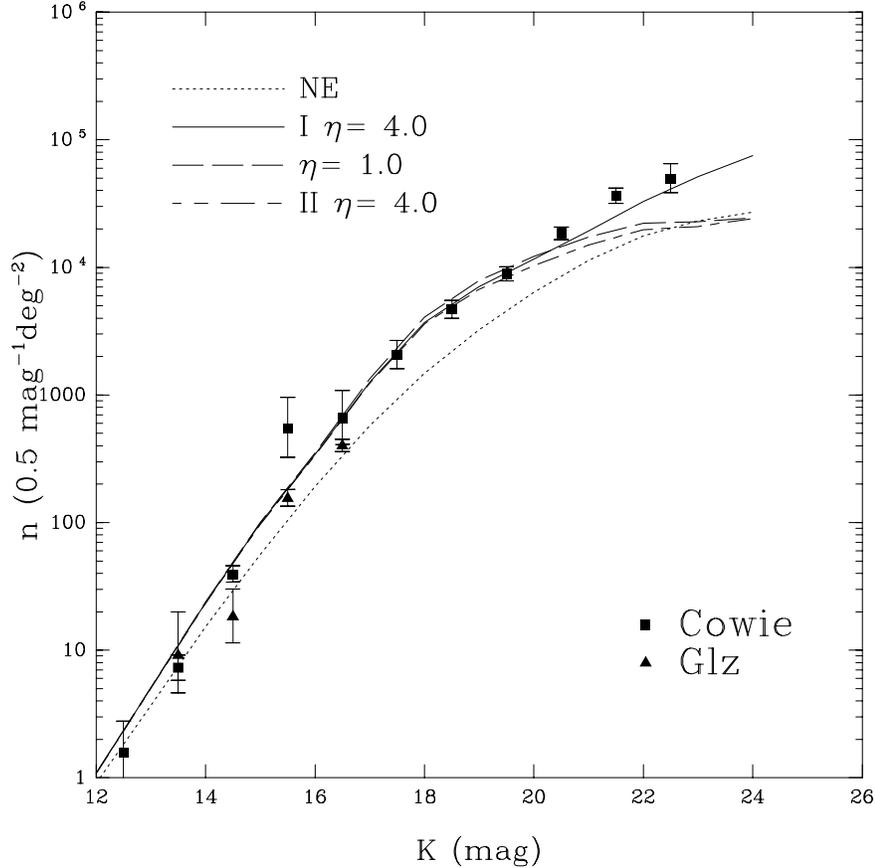}
        \caption{Number of galaxies against \K\ magnitude for the
same models as Figure 3, except that the NE KGB model has been
replaced by one whose contribution to $L^*$ in the \K-band due to
starbursts is neglected (II $\eta = 3.0$).}
\end{figure}

\begin{figure}
        \vspace{12cm}
        \includegraphics{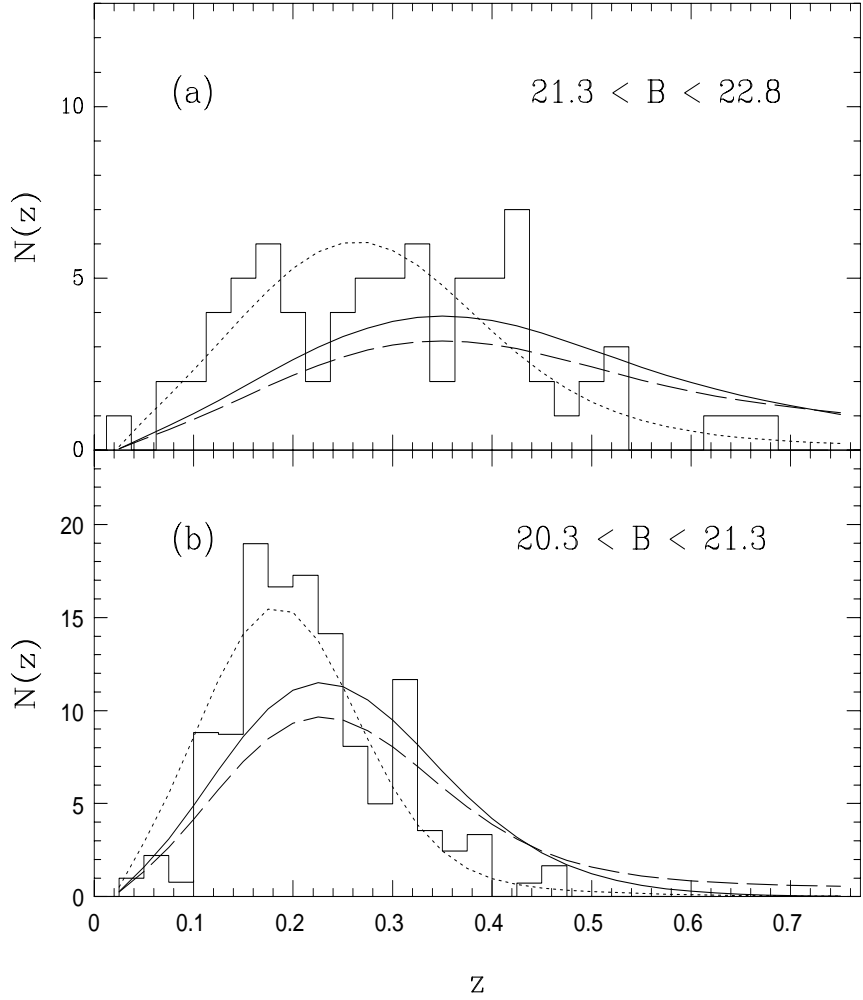}
\caption{Redshift distributions in two intervals of \B\ magnitude:
(a) $20.0 \le b_J \le 21.0$, and (b) $21.0 \le b_J \le 22.5$. Data
are from Broadhurst et al. (1988) and
Colless et al. (1990; 1993). The approximate color relationships
$B = b_J + 0.18$ and $B = b_J + 0.16(b_J-r_F)$ for Broadhurst et al.
and Colless et al. data, has been used, respectively. The two
merger-driven and our NE models are plotted. The line code is the
same as in Fig. 3.}
\end{figure}

\section{CONCLUSIONS}

Differences of up to 0.25 mag in V-K are found when various
photometric calibrations are used. On the other hand, a
smaller difference, 0.1 mag in V-K, is found when
we compare the results that come from using two
reasonable IMFs, the so-called KTG IMF and the Salpeter one.
This is certainly negligible as far as the B or K number distributions
calculations are concerned.

By using the best parameters we have in the literature for the
LFs of the different classes of galaxies and for the mix of them,
we arrive to the conclusion that
number and luminosity evolution of the galaxy luminosity function
are important ingredients in the explanation of the excess observed
in the \B-counts.
Our merger-driven models require little evolution both in number
and luminosity;
for example, at $z = 0.4$ the total number of galaxies has just
increased by 14 \% and the percentage of I galaxies is just 17 \% (model
with $\eta= 4.0$). Moreover, luminosity evolution is such that
a model with $\eta= 4.0$ predicts that about 13 \% of galaxies
have $z > 0.7$ in the $21.0 < m_{b_J} < 22.5$ range.
Models with more emphasis in number evolution reproduce
better the redshift distributions (the $\eta = 4.0$ model
as compared with the $\eta = 1.0$ one), as it should be expected.
Despite our relative success of our models,
the excess of high redshift of galaxies predicted by
them indicates us that number and luminosity
evolution may not be the whole story.

\vspace{1.0cm}
\centerline{\bf ACKNOWLEDGMENTS}
\vspace{0.5cm}
I would like to thank Manuel Peimbert and Deborah Dultzin for their careful
reading of the manuscript. I am grateful to the anonymous referee for
his/her comments on the manuscript.
This paper was supported in part by Direcci\'on
General del Personal Acad\'emico (DGAPA) through grant IN-1000994 at
Universidad Nacional Aut\'onoma de M\'exico.

\begin{center}
\noindent {\LARGE \bf References}
\end{center}

\ref Babul A., \& Rees, M.J. 1992, MNRAS, 255, 346
\ref Bessell, M. S., \& Brett, J. M. 1988, PASP, 100, 1134,
\ref Broadhurst, T. J., Ellis, R. S., \& Shanks, T. 1988, MNRAS, 235, 827
\ref Broadhurst, T. J., Ellis, R. S., \& Glazebrook, K. 1992, Nature, 355, 55
\ref Bruzual, A. G. 1983, ApJ, 273, 105
\ref Bruzual, A. G. 1992, in Cosmology and Large-Scale Structure in the
Universe ed. R. R. de Carvalho (ASP, 24), 43
\ref Bruzual, A. G., \& Charlot, S. 1993, ApJ, 405, 538
\ref Carlberg, R. G., \& Charlot, S. 1992, ApJ, 397, 5
\ref Charlot, S., \& Bruzual, A. G. 1991, ApJ, 367, 126
\ref Cole, Shaun, Ellis, R. S., Broadhurst, T. J., \& Colless, M. M.
1994, MNRAS, 267, 541
\ref Col\'in, P., Schramm, D. N., \& Peimbert, M. 1994, ApJ, 426, 459 (CSP)
\ref Coleman, G. D., Wu, C. -C., \& Weedman, D. W. 1980, ApJS, 43, 393 (CWW)
\ref Colless, M. M., Ellis, R. S., Taylor, K., \& Hook, R. N. 1990,
MNRAS, 244. 408
\ref Colless, M. M., Ellis, R. S., Broadhurst, T. J., Taylor, K., \&
Peterson, B. A. 1993, MNRAS, 261, 19
\ref Colless, M., Schade, D., Broadhurst, T. J., \& Ellis, R. S. 1994, MNRAS,
in press
\ref Cowie, L. L., Songaila, A., \& Hu, E. M. 1991, Nature, 354, 460
\ref Cowie, L. L., Gardner, J. P., Wainscoat, R. J., \& Hodapp, K. W.
1993, ApJ, submitted
\ref de Lapparent, V., Geller, M. J., \& Huchra, J. P. 1989, ApJ, 343, 1
\ref Eales, S. 1993, ApJ, 404, 51
\ref Efstathiou, G., Ellis, R. S., \& Peterson, B. A. 1988, MNRAS, 232, 431
\ref Efstathiou, G., Bernstein, G., Katz, N., Tyson, J. A., \&
Guhathakurta, P. 1991, ApJ, 380, L47
\ref Ferguson, H. C., \& Sandage, A. 1991, AJ, 101, 765
\ref Fukugita, M., Takahara, F., Yamashita, K., \& Yoshii, Y. 1990, ApJ, 361,
L1
\ref Gardner, J. P., Cowie, L. L., \& Wainscoat, R. J. 1993, ApJ, L9
\ref Griffiths, R. E., et al. 1994a, ApJ, 437, 67
\ref Griffiths, R. E., et al. 1994b, ApJ, 435, L19
\ref Guiderdoni, B., \&  Rocca-Volmerange, B. 1987, A\&A, 186, 1
\ref Infante, L., \& Pritchet, C. J. 1994, preprint
\ref Johnson, H. L. 1966, ARA\&A, 4, 193
\ref Kroupa, P., Tout, C. A. \& Gilmore, G., 1993, MNRAS, 262, 545
\ref Koo, D. C., Gronwall, C., \& Bruzual, G. A. 1993, ApJ, 415, L21 (KGB)
\ref Lilly, S. J., Cowie, L. L., \& Gardner, J. P. 1991, ApJ, 369, 79
\ref Lonsdale, C. J., \& Chokshi, A. 1993, AJ, 105, 1333
\ref Loveday, J., Peterson, B. A., Efstathiou, G., \& Maddox, S. J.
1992, ApJ, 390, 338
\ref Maddox, S. J., Sutherland, W. J., Efstathiou, G., Loveday,
J., \& Peterson, B. A. 1990, MNRAS, 247, 1P
\ref McGaugh, S. S. 1994, Nature, 367, 538
\ref Metcalfe, N., Shanks, T., Fong, R., \& Jones, L. R. 1991,
MNRAS, 249, 498
\ref Peimbert, M., Sarmiento, A., \& Col\'in, P. 1994,
RevMexAA, 28, 181
\ref Pence, W. 1976, ApJ, 203, 39
\ref Pilyugin, L. S. 1993, A\&A, 94, 175
\ref Sandage, A., Binggeli, B., \& Tammann, G. A. 1985, AJ, 90, 1759
\ref Schaller, G., Schaerer, D., Meynet, G., \& Maeder, A. 1992, A\&AS,
96, 269
\ref Schmidt-Kaler, T. H. 1982, in
Landolt-Bornstein New Series, Vol. 2b, A\&A -stars and star clusters-
eds. K. Schaifers \& H.H. Voigt (New York:Springer-Verlag)
\ref Soifer, B. T., et al. 1994, ApJ, 420, L1
\ref Tyson, J. A. 1988, AJ, 96, 1
\ref Worthey, G. 1994, ApJS, submitted
\ref Yoshii, Y. 1993, ApJ, 403, 552
\ref Yoshii, Y., \& Takahara, F. 1988, ApJ, 326, 1
\ref Zucca, E., Pozzetti, L., \& Zamorani, G. 1994, MNRAS, in press

\end{document}